# DGSSA: Domain generalization with structural and stylistic augmentation for retinal vessel segmentation

Bo Liu[1], Yudong Zhang[2], Shuihua Wang[3], Siyue Li[4], Jin Hong[1,*]

1. School of Information Engineering, Nanchang University, Nanchang, 330031, China
2. School of Computer Science and Engineering, Southeast University, Nanjing, 210096, China
3. Department of Biological Sciences, School of Science, Xi'an Jiaotong Liverpool University, Suzhou, 215123, China
4. Department of Radiological Sciences, University of California Los Angeles, Los Angeles, CA, United States

E-mail: liuboncu@email.ncu.edu.cn; yudongzhang@ieee.org; Shuihua.Wang@xjtlu.edu.cn; siyueli@mednet.ucla.edu; hongjin@ncu.edu.cn;
* Correspondence should be addressed to Jin Hong

**Abstract:** Retinal vascular morphology plays a crucial role in diagnosing diseases such as diabetes, glaucoma, and hypertension, making accurate segmentation of retinal vessels essential for early intervention. Traditional segmentation methods assume that training and testing data share similar distributions, which can lead to poor performance on unseen domains due to domain shifts caused by variations in imaging devices and patient demographics. This paper presents a novel approach, DGSSA, for retinal vessel image segmentation that enhances model generalization by combining structural and stylistic augmentation strategies. We utilize a space colonization algorithm to generate diverse vascular-like structures that closely mimic actual retinal vessels, which are then used to generate pseudo-retinal images with an improved Pix2Pix model, allowing the segmentation model to learn a broader range of structure distributions. Additionally, we utilize PixMix to apply random photometric augmentations and introduce uncertainty perturbations, enriching the stylistic diversity of fundus images and further improving the model's robustness and generalization across varying imaging conditions. Our framework, which employs a DeepLabv3+ model with a MobileNetV2 backbone as its segmentation network, has been rigorously evaluated on four challenging datasets—DRIVE, CHASEDB1, HRF, and STARE—achieving Dice Similarity Coefficient (DSC) of 78.45%, 78.62%, 72.66% and 82.17%, respectively, with an average DSC of 77.98%. These results demonstrate that our method surpasses existing approaches, validating its effectiveness and highlighting its potential for clinical application in automated retinal vessel analysis.
**Keywords:** Medical image segmentation, domain generalization, data augmentation, structural augmentation

## 1. Introduction

Morphological changes in retinal vasculature are closely associated with various diseases, such as diabetes, glaucoma, and hypertension [1] , retinal vessel segmentation can assist clinicians in the early diagnosis and quantification of vascular abnormalities. However, in clinical practice, manually annotating all minute vessels in fundus images by expert ophthalmologists is both subjective and labor-intensive. Therefore, the development of an automated retinal vessel segmentation algorithm holds significant clinical value [2] . Recently, deep learning have been applied to automated segmentation tasks, including optic disc/cup (OD/OC), retinal vessels, and lesions, achieving remarkable performance [3, 4] . Most methods typically assume that the training and testing images share similar distributions assumed aligned domain generalization. In clinical scenarios, this assumption often does not hold due to various factors, including differences in scanning devices, imaging protocols, patient demographics, and natural distribution variations between training and testing data. These discrepancies, referred to as domain shifts [5], can significantly degrade model performance during inference stage [6, 7] . Wang et al [8] demonstrated that the segmentation performance of Convolutional Neural Networks (CNNs) rapidly declines on new datasets from unknown domains, particularly when there is a huge distributional discrepancy between the training dataset and the new data. Given that it is impractical to preemptively collect all possible types of datasets for training in order to cover distribution of unseen data [9] , developing a robust CNN-based retinal image segmentation algorithm capable of generalizing to unseen datasets poses an urgent and challenging problem in the field.

Improving the generalization ability of models on new datasets is an eternal theme in machine learning and deep learning, and many research works have been devoted to solving this problem from the past to the present [10] . Among them, transfer learning is one of the most widely adopted approaches to mitigate dataset bias and domain shifts [11] . It involves pre-training a model on a source domain and fine-tuning it on a different but related target domain. However, a key limitation of transfer learning is its reliance on labeled images from the target domain. To alleviate the dependency on costly data annotation, unsupervised domain adaptation (UDA) has emerged as a promising solution [12] . In recent years, it has increasingly played a significant role in medical artificial intelligence, particularly in the field of medical image segmentation, where various effective UDA methods have been developed [13-15]. UDA seeks to adapt models to new target domains without the need for annotated labels, it still requires access to data from the target domain. Moreover, UDA often necessitates fine-tuning the model to adapt to the new domain, which can be infeasible in clinical practice [16] . Hence domain generalization (DG)for unseen domains has garnered increasing attention as an active area of research with promising potential in image analysis [17] . DG aims to develop a model that learns from diverse training datasets and generalizes well to any unseen target domain, eliminating the need for target domain data collection or model fine-tuning. In the field of medical image analysis, DG has demonstrated considerable impact in clinical workflows where time and resources are constrained, as it delivers robust model performance across diverse and unseen datasets without the need for additional domain-specific interventions.

In recent years, Domain Generalization (DG) has shown significant promise in medical image segmentation, particularly in retinal vessel segmentation [18], with numerous high-quality studies emerging in the field [19-21] , While these studies highlight advancements in segmentation accuracy, their practical applicability in clinical settings remains limited, indicating substantial room for improvement. Current methods primarily focus on diversifying image styles to enable models to learn semantic features independent of style variations, thereby facilitating accurate segmentation of unseen data. This approach assumes that the structure distribution of unseen domains aligns with that of the training domain, mainly addressing visual factors such as color, texture, and contrast variations across datasets. However, significant anatomical and physiological differences in retinal vascular structures exist among different populations and ethnicities, driven by factors such as genetic background, geographical location, age, and health status. These differences can lead to variations in vessel morphology, density, branching patterns, and the ratio of arterioles to venules. For instance, studies have shown that retinal arterioles in African Americans are generally narrower than those in Caucasians, and aging significantly affects the diameter and curvature of retinal arterioles. Additionally, genetic predispositions, such as susceptibility to diabetes and hypertension, contribute to further variability in retinal vascular morphology within the same population [22-24]. Given these findings, we argue that methods addressing only style distribution discrepancies between training and unseen data are insufficient for overcoming the generalization challenges in retinal vessel segmentation. Structure distribution bias plays a crucial role in limiting the model's generalization capability. When trained on a dataset that captures the retinal structures of a specific population, this bias can significantly impair segmentation performance on unseen domains, particularly those with markedly different vascular morphologies.

To address this challenge, we propose a novel and effective method for generalizing retinal vessel image segmentation that combines structural and style augmentation strategies to enhance the generalization capability of segmentation networks. For structural augmentation, we utilize a spatial colonization algorithm specifically designed for retinal vessel segmentation to generate artificial yet highly realistic vascular-like structures that replicate the branching and geometric features of actual retinal vessels. By integrating this approach with an improved Pix2Pix model, we can produce a variety of pseudo-retinal images featuring diverse vascular shapes and branching structures. This diversity allows the segmentation model to learn a broader range of structure distributions, reducing reliance on specific vascular structural features present in the training data. For style augmentation, we employ PixMix to apply random photometric enhancements and introduce uncertainty perturbations, enriching the stylistic diversity of fundus vessel images. This diversity enhances the model's ability to adapt to different imaging conditions, such as variations in lighting and contrast across datasets. By simultaneously enhancing both structural and style diversity, our method has achieved satisfactory performance. The main contributions of this paper are as follows:

(i) We uniquely combine structural augmentation with widely used style augmentation techniques for the first time, proposing

a simple and effective domain generalization framework for fundus retinal segmentation. This approach significantly expands sample diversity and improves generalization.

(ii) We employ a spatial colonization algorithm to randomly generate diverse vascular-like structures that closely resemble retinal blood vessels. These structures are then processed using an enhanced Pix2Pix model with multi-scale discriminators, resulting in retinal images that feature a wide variety of intricate vascular patterns. This diversity in the generated images enables the segmentation model to learn a broader range of retinal structures during training.

(iii) We utilize PixMix to apply random photometric augmentations and introduce uncertainty perturbations, enriching the stylistic diversity of fundus vessel images, which further enhances the model's robustness and generalization across varying conditions.

(iv) We rigorously evaluate our framework on four challenging retinal vessel structure segmentation datasets—DRIVE, CHASEDB, HRF, and STARE—demonstrating state-of-the-art (SOTA) performance that surpasses existing methods, thereby validating the effectiveness and potential of our proposed approach.

## 2. Related work

### 2.1. Retinal vessel Segmentation

Early research on retinal vessel segmentation primarily relied on traditional image processing techniques [25-27], such as handcrafted filters and morphological operations, which generally delivered suboptimal performance. In recent years, deep learning-based methods have achieved significant advancements, with U-Net [28] being one of the most widely used models for medical image segmentation. Despite its remarkable accuracy compared to traditional approaches, U-Net still struggles to effectively segment the complex retinal vasculature. To mitigate the loss of space information caused by the successive pooling operations in U-Net, CE-Net [29] introduced dilated dense blocks and residual pooling. U-Net++ [30] redesigned the skip connections within U-Net to enhance multi-scale information fusion. In CS-Net [31], space attention and channel attention mechanisms were employed to facilitate the integration of local and global information, aiding in the accurate capture of vascular morphology. A Review for retinal vessel segmentation [32] investigated various segmentation frameworks and explored the feasibility and limitations of applying deep learning to fundus vessel segmentation. However, most of these methods focus on improving segmentation performance on known datasets through better network architectures or leveraging prior knowledge. They tend to overfit to specific data distributions and fail to generalize to unseen target datasets [33] assumed aligned domain generalization. To eliminate the impact of network architecture on generalization performance, we adopted the same segmentation framework as prior state-of-the-art methods—DeepLabv3+ with a MobileNetv2 backbone—as our segmentation model. This segmentation model corresponds to the component labeled "S" in Figure 1.

### 2.2. Data Augmentation

Robustness from the perspective of data augmentation has been extensively explored. Current mainstream methods predominantly focus on style enhancement to improve generalization capabilities. For instance, AugMix and its variants [34, 35] proposed diversifying the training distribution by applying randomly sampled augmentations in a cascaded manner. DeepAugment [36] introduced the use of image translation models to generate new training images. Additionally, Autoaugment [37] proposed a custom augmentation strategy designed to improve robust generalization. Various methods, such as ensembles of expert models tuned to specific frequencies [38], maximum entropy augmentations [39], spectral perturbations [40], and fractal and feature map-based augmentations [41], have proven successful in tackling the challenges posed by distribution shifts. However, existing data augmentation approaches primarily concentrate on style diversification, modifying the visual appearance of images while preserving their underlying structural integrity. In the context of medical images, there are significant differences in the structure of the segmented target object, making it crucial to explore both style enhancement and structural enhancement. This combined approach aims to enrich the dataset's visual variability while ensuring that the critical structural features inherent in medical images are adequately represented.

## 2.3. Medical Image Synthesis

Generative Adversarial Networks (GANs) [42] have become a cornerstone in medical image synthesis, with significant applications in intra-modality augmentation [43] , cross-domain image-to-image translation [44] , quality enhancement [45] , and missing modality generation [46] . In this context, we briefly review prior work on retinal image synthesis, which closely relates to our research. Costa et al. [47] employed U-Net with a conditional GAN, Pix2Pix [48] , to map vascular masks to corresponding fundus images. To simplify their framework, they proposed an adversarial autoencoder (AAE) for vascular synthesis alongside a GAN for generating retinal images [49].

Similarly, Guibas et al. [50] introduced a two-stage approach using a DCGAN to generate vasculature from noise and a cGAN (Pix2Pix) to synthesize fundus images, though this requires paired images and masks. These methods often rely on additional vascular annotations, resulting in morphologically unrealistic vessels and a lack of diversity. To address these limitations, Zhao et al. [51] developed Tub-sGAN, integrating style transfer to produce diverse outputs, while SkrGAN [52] introduced sketch-based prior constraints. Lin et al. [53] proposed providing GANs with pre-extracted real backgrounds, which establishes an implicit skip-connection mechanism. This approach enables the GAN to concentrate its learning capacity on mapping foreground regions while leveraging preserved background information, thereby significantly enhancing the generation quality of fundus vasculature in retinal images. Building on these advancements, we propose a novel approach using a space colonization algorithm for vascular-like structures and synthesize retinal images via an improved Pix2Pix. To enhance stability and accuracy, we design a multi-scale discriminator that trains on both unpaired generated vascular structures and paired ones, allowing for the production of realistic and diverse retinal images.

## 2.4. Domain Generalization in Medical Image Segmentation

Significant research interest has emerged in domain generalization (DG) for medical image segmentation, particularly as traditional image transformations demonstrate effectiveness in mitigating domain shifts across diverse medical datasets. For instance, Zhao et al. [54] leveraged shape and intensity information from unannotated MRI images to deform and modify atlases, achieving substantial advancements in one-shot brain MRI segmentation methods. Meanwhile, Chen et al. [55] meticulously designed data normalization and augmentation strategies, enhancing the generalization of CNN-based models in cardiac MR image segmentation tasks. Otarola et al. [56] validated the superiority of traditional image transformations in computational pathology, while BigAug [57] expanded these techniques by stacking additional augmentations, facilitating better generalization to unseen domains. Liu et al. [20] introduced the concept of continuous frequency space (ELCFS) for scenario learning, effectively bridging distribution gaps among multiple source domains. Representation learning has also gained traction; Wang et al. [19] proposed the Domain Feature Embedding (DoFE) framework, which integrates domain priors and an attention mechanism to fuse memorized domain features with image features, achieving robust retinal image segmentation. Additionally, AADG [21] introduces an innovative module for style enhancement that leverages the Sinkhorn distance in unit sphere space to maximize the diversity of styles across multiple augmented domains. This approach facilitates the automatic enhancement process, making it more effective in managing variations in style. Based on these basic works, we further enhanced domain generalization by combining structure and style enhancement, significantly improving the robustness and performance of retinal vessel segmentation models in various medical image analysis environments.

## 3. Method

Domain Generalization (DG) seeks to create models that perform well on unseen domains by learning from multiple source domains with distinct data distributions. In retinal vessel segmentation, the challenge lies in the variability of both structural and style distributions across populations, devices, and imaging conditions. Let $X$ denote the input space (e.g., retinal images) and $Y$ the output space (e.g., vessel segmentations). A domain is composed of data that are sampled from a joint distribution of the input sample and output label $P_{xy}$ , We represent a domain $T = (x_j, y_j)_{j=1}^{n} \sim P_{xy}$ , where $x \in X$ and $y \in Y$ , and $n$ is the number of samples. In the DG setting, we have access to $m$ source domains $T_{source} = \{T^i \mid i = 1, \ldots, m\}$, where each domain $T^i = (x_j^i, y_j^i)_{j=1}^{n_i}$ donates the $i$-th domain with $n_i$ data pairs. These distributions differ across domains, reflecting both style and structural variability in the data $P_{XY}^i \neq P_{XY}^j, 1 \leq i \neq j \leq m$. The objective of DG is to learn a predictive model $h_\theta: X \to Y$, which generalizes well to unseen target domains $T_{target} = \{T^i \mid i = m+1\}$, where $T_{target}$ cannot be accessed during training. Formally, we aim to minimize the generalization error:

$$E_{(x,y) \sim T_{target}}[L(h_\theta, y)] \tag{1}$$

where $L$ is a suitable loss function, such as the cross-entropy loss. To mitigate the domain shifts between the source and target domains, we introduce a structure-enhanced learning framework named Structure-Style-Augment framework.

## 3.1. The Struct-Style-Augment Framework

The Structure-Style-Augment framework, as illustrated in Figure 1, integrates structural and stylistic enhancement strategies to augment the diversity of synthesized retinal images, thereby enhancing generalization capabilities. This framework is divided into three steps. The first step employs the Space Colonization Algorithm [58] to generate vessel-like structures, simulating the growth of branching networks or tree-like architectures. This approach, combined with additional processing, effectively captures the intricate patterns associated with vascular formation and spatial organization features. In the second step, we employed an enhanced Pix2Pix model with a multi-scale discriminator to independently train on each dataset. Here, the generated masks were utilized alongside the original masks of paired retinal images, enabling the generation of retinal images characterized by various complex vascular patterns. Finally, the third step integrated the application of PixMix to implement random photometric augmentation, introducing uncertainty perturbations to enhance style. Throughout this process, both the structurally enhanced dataset and the original dataset were used for style augmentation. These were subsequently combined with the segmentation model and trained across multiple source domains, thereby achieving effective segmentation performance in the target domain.

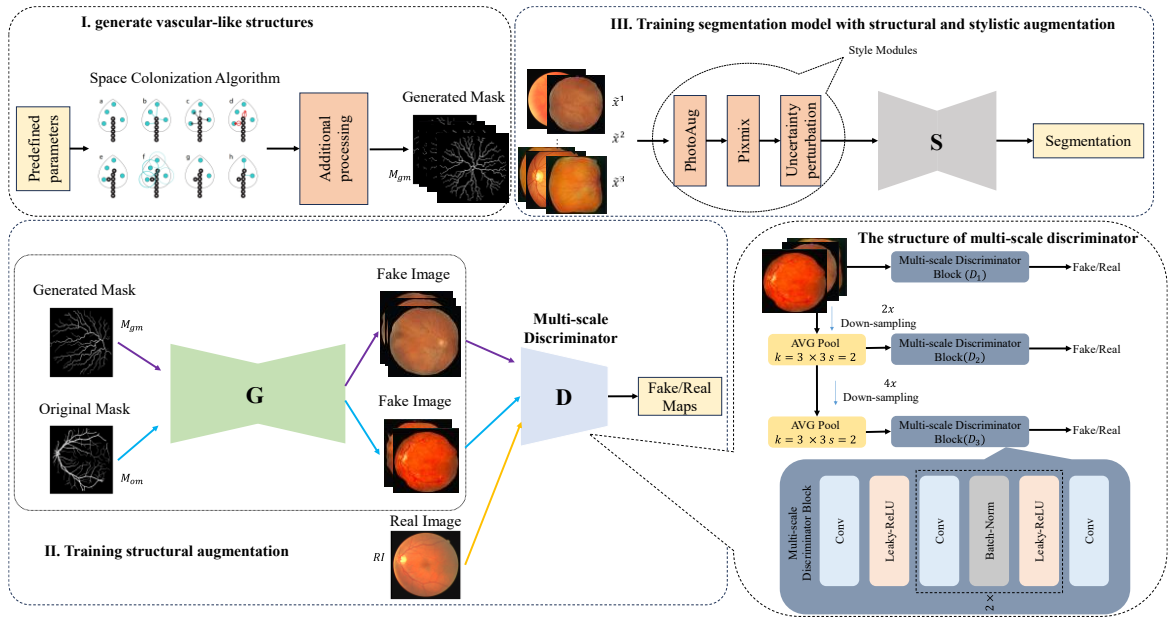

Figure 1 The Struct-Style-Augment framework encompasses three primary components: a generator designed to produce curved structural patterns, a training module focused on generating appropriate pseudo-retinal vessel samples, and a suite of style enhancement modules that collaboratively train the segmentation model. Additionally, the framework features a multi-scale discriminator architecture, illustrated in the bottom right corner, which enhances the overall performance of the system.

## 3.2. Structural Augmentation

The Space Colonization Algorithm, a procedural modeling technique in computer graphics, simulates the growth of branching networks or tree-like structures [58], including vascular systems, leaf venation, and root systems. This algorithm models the iterative growth of curve structures characterized by two fundamental elements: attractors and nodes. Space colonization is a process that iteratively grows branching networks based on the spatial distribution of growth hormone sources (called "auxin" sources), which serve as attractors in the model. It treats spatial competition as the critical factor determining branch placement and density, ensuring that regions with dense attractor clusters develop finer, more intricate branching patterns. The core steps of this process, illustrated in the upper left of Figure 1, are as follows. First, the tree structure is initialized with six nodes (black discs with white centers) and four attractors (blue discs). The steps are as follows: (a) place a set of attractors; (b) identify which attractors influence which nodes — each attractor is associated with its nearest tree node, provided that the node lies within a predefined attraction distance (represented by the blue lines in the figure); (c) for each node, calculate the average direction of all influencing attractors; (d) determine the position of new nodes by normalizing this average direction to a unit vector and scaling it by a predefined segment length; (e) place the nodes at the computed positions; (f) check if any nodes fall within the kill zone of any attractors; (g) remove attractors that do not meet the desired growth criteria; (h) restart the process from step (b) until the maximum number of nodes is reached. Additionally, to simulate venous thickening and represent the varying diameters of blood vessels, a method has been proposed that incorporates the thickness of each node into the thickness of its parent node. Specifically, this is expressed in terms of the radii $R$, $R_1$, $R_2$, denoting the radius of the parent branch and its two child branches, respectively. In this context, $n$ is a parameter of the method, typically set to 3 in accordance with Murray's law.

$$R^n = R_1^n + R_2^n \tag{2}$$

We begin with several thousand attractor points and a single initial root node. At each iteration, each attractor is associated with its nearest node within a predefined attraction radius $d$; if the distance is less than $d$, that attractor contributes to the node's growth direction. If any node enters the attractor's kill radius, the attractor is immediately pruned. In our implementation, the predefined parameters are set as follows: the attraction radius is 5 units, the segment length is 20 units, and the kill radius is 5 units.

To further refine the vascular structures, we incorporate random perturbations in attractor positions. We then apply post-processing techniques such as thresholding to eliminate spurious offshoots, retaining the largest connected component to ensure vascular coherence. Morphological erosion is also applied to thin the primary vessels, thereby enhancing the fidelity and increasing the number of finer branches. Finally, the generated curve structures are adapted to the region of interest (ROI) of each dataset, ensuring that the curves do not extend beyond the corresponding ROI boundaries, thereby forming the final vascular structure map.

---

**Algorithm 1** Vascular Structure Generation via Space Colonization Algorithm

**Input:** Set of attractors $A = \{a_1, a_2, \ldots, a_n\}$, Initial root node $n_0$, Attraction radius $d$, Kill radius $k$, Segment length $l$, Maximum node count $M$
**Output:** Final vascular structure $\mathcal{T}$

1. **Initialize:**
2.     Create initial tree with $\mathcal{T}$ root node $n_0$, Randomly perturb attractor positions in $A$
3. **Repeat until number of nodes in $\mathcal{T} \geq M$:**
4.     **(a) Associate attractors:**
5.         For each attractor $a_i \in A$,
6.             Find nearest node $n_j \in \mathcal{T}$ such that $\|a_i - n_j\| < d$
7.             Associate $a_i$ with $n_j$
8.     **(b) Compute growth directions:**
9.         For each node $n_j$ with associated attractors,
10.             Compute average growth direction vector $\vec{v_j}$
11.             Normalize $\vec{v_j}$, scale by segment length $l$
12.     **(c) Add new nodes:**
13.         For each node $n_j$,
14.             Create new node $n_{j+1} = n_j + l \cdot \vec{v_j}$
15.             Add $n_{j+1}$ to $\mathcal{T}$

16. **(d) Prune attractors:**
17.     For each attractor $a_i \in A$,
18.         If any node $n \in \mathcal{T}$ satisfies $\|n - a_i\| < k$,
19.             Remove $a_i$ from $A$
20. **Post-processing:**
21.     Apply thresholding to remove spurious branches
22.     Retain the largest connected component
23.     Apply morphological erosion to refine vessel thickness
24. **Fit to ROI:**
25.     Trim final structure to fit within the region of interest (ROI) of dataset
26. **Return:** Final vascular tree structure $\mathcal{T}$

---

Algorithm 1 outlines the procedural steps of the space colonization algorithm tailored for vascular structure generation. To further illustrate this process, Figure 2 provides a visual demonstration of the generation of our vascular structure.

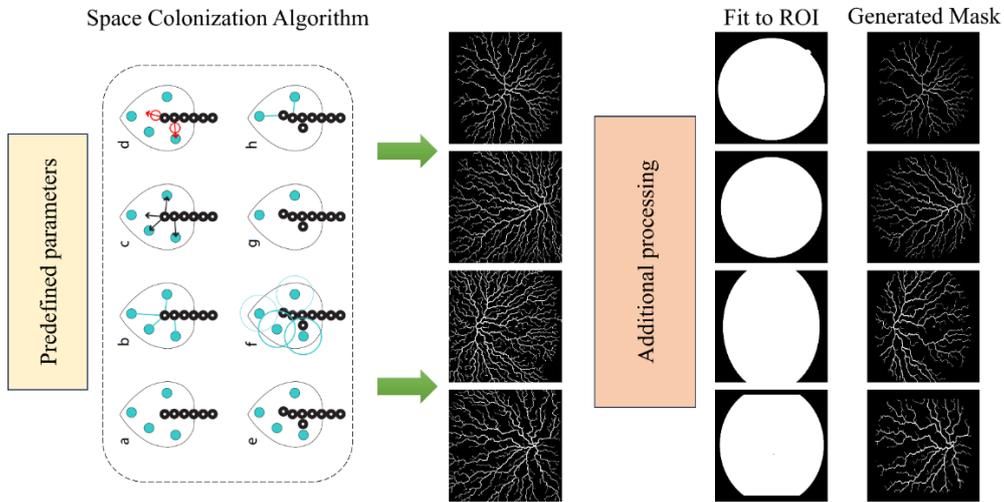

Figure 2 This process generates vascular-like structures that are designed in accordance with the region of interest (ROI) specific to each dataset.

To synthesize more realistic retinal images, we employed an improved Pix2Pix method, as illustrated in Figure 1. The generator(G) is structured as a U-Net architecture, tasked with synthesizing pseudo-retinal vessel images from the generated mask $M_{gm}$ and the original mask $M_{om}$. The discriminator(D) features a multi-scale architecture (as shown in the lower right corner of Figure 1, comprising three sub-modules at different scales, all sharing the same network structure. Each convolutional block within these sub-modules consists of a convolutional layer, a batch normalization (BN) layer, and a Leaky ReLU activation function. The number of filters in the convolutional layers increases progressively, starting with $n = 128$ in the first layer and $n = 256$ in the second layer. Within the discriminator architecture, all convolutional layers utilize $4 \times 4$ kernels, with the first convolutional layer and the convolutional blocks applying a stride of 2, while the final convolutional layer employs a stride of 1. The multi-scale discriminator consists of several independent discriminators $D_s$ where $s = \{1, 2, ..., S\}$ denotes the scale. Each discriminator operates on images of different resolutions; for instance, discriminator $D_1$ processes images at the original resolution, while $D_2$ and $D_3$ handle images that have been down-sampled by factors of 2 and 4, respectively. The goal of the discriminator is to ensure that the generated images possess realistic details at multiple levels of supervision. During the discriminator training phase, the generator utilizes feedback from the multi-scale discriminator, enabling it to improve the generated images at varying scales. For example, the low-resolution discriminator $D_3$ assists the generator in capturing the global vascular morphology, while the high-resolution discriminator $D_1$ focuses on subtle local details, such as vessel edges and finer branches. Here, we compute the parallel mean of the three discriminators across different scales and sum the results to obtain $D$. This approach effectively integrates information from multiple scales, enhancing the discriminator's ability to undergo comprehensive training. Furthermore, to enhance the generation of realistic retinal images, we modify the training process to include the generated vessel structures in the generator's training cycle, rather than solely relying on paired data as in the original Pix2Pix methodology, and subsequently feeding these generated retinal images into the discriminator for further training. Consequently, we have further refined the design of the loss function to facilitate comprehensive training on both paired and unpaired data. The formulation of our loss function is as follows. In our generator loss, we incorporate both adversarial loss $L_{adv}$ and consistency loss $L_{L1}$, formulated as follows:

$$L_{L1} = \mathbb{E}_{M_{om}, RI}[\| G(M_{om}) - RI \|_1] \tag{3}$$

$$L_{adv} = \mathbb{E}_M [logD(G(M))] \tag{4}$$

where $RI$ is the real images and M is the masks.

Given that unpaired data lacks corresponding retinal vessel images, we train only the adversarial loss for this data. Thus, the total generator loss $L_G$ is expressed as:

$$L_{G_{paired}} = \lambda_{L1} L_{L1} + \lambda_{adv} \mathbb{E}_{om} [logD(G(M_{om}))] \tag{5}$$

$$L_{G_{unpaired}} = \lambda_{adv} \mathbb{E}_{gm} [logD(G(M_{gm}))] \tag{6}$$

$$L_G = L_{G_{paired}} + L_{G_{unpaired}} \tag{7}$$

where $L_{G_{paired}}$ is the generator loss of paired data and $L_{G_{unpaired}}$ is the generator loss of unpaired data.

In the discriminator's loss, we introduce gradient penalty (GP) to enhance the effectiveness of adversarial training. The loss functions are defined as follows:

$$L_{D_{real}} = -\mathbb{E}_{RI} [logD(RI)] \tag{8}$$

$$L_{D_{fake_{paired}}} = \mathbb{E}_{M_{om}} \left[\log\left(1 - D(G(M_{om}))\right)\right] \tag{9}$$

$$L_{D_{fake_{unpaired}}} = -\mathbb{E}_{M_{gm}} \left[\log\left(1 - D(G(M_{gm}))\right)\right] \tag{10}$$

where $L_{D_{real}}$ represents the discriminator loss associated with real images, $L_{D_{fake_{paired}}}$ denotes the discriminator loss for fake images that correspond to paired masks, and $L_{D_{fake_{unpaired}}}$ refers to the discriminator loss for fake images that lack paired masks.

Therefore, the overall loss function for the discriminator $L_D$ is given by:

$$L_D = \lambda_1 \left(L_{D_{real}} + L_{D_{fake_{paired}}} + L_{D_{fake_{unpaired}}}\right) + \lambda_{GP} \cdot (GP_{paired} + GP_{unpaired}) \tag{11}$$

where $GP_{paired}$ is the gradient penalty for paired images and $GP_{unpaired}$ is the gradient penalty for unpaired images.

### 3.3. Style Augmentation

We propose an enhanced data augmentation technique, PixMix, PhotoAug and Uncertainty Perturbation (UP), to increase model robustness by combining external images, probabilistic perturbations, and controlled mixing operations. The results are illustrated in Figure 3. As shown in the figure, we extracted a subset of our training dataset, which includes both structurally enhanced samples and original data. Randomized style augmentations were applied to both subsets to simulate diverse visual appearances and improve model generalization. The augmentation process introduces randomness in both the choice of stylistic transformations and the use of uncertainty-driven perturbations, resulting in a wide range of image variations. Notably, the style-enhanced synthetic images (Rows Figure 3b and Figure 3d) do not necessarily exhibit higher contrast or sharper structures than the original images (Rows Figure 3a and Figure 3c), which is intentional. Instead of pursuing visually clearer samples, our objective is to generate more challenging cases to encourage the model to learn from harder examples and thus improve its generalization performance. Given an original input image $x \in X$ and an external image $z \in Z$ drawn from a set of auxiliary images, the process begins with a random initialization step where an initial transformation is applied to $x$. This transformation is chosen from a set of augmentations (denoted as PhotoAug) or left as the original image. Here, $\tilde{x}_0$ serves as the starting point for subsequent mixing operations. The number of mixing rounds $K$ is predefined, and for each round $k$, a random number of mixing steps $T$ is selected, where $T \in \{0, 1, ..., K\}$. Within each mixing step $t$, an image $x_{mix}$ is sampled either from the set of augmentations applied to $x$ or from the external image $z$. Following the selection of the image $x_{mix}$, a mixing operation $mix\_op$ is chosen randomly from a set of element-wise operations such as addition or multiplication. This introduces variability by altering the combination of the input image and the auxiliary data. Additionally, at each step, the method incorporates distributional uncertainty by applying perturbations based on statistical measures derived from the mixed image. Specifically, if a randomly drawn probability is less than a pre-defined threshold $p$, the mean $\mu$ and standard deviation $\sigma$ of the selected image $x_{mix}$ are computed. Here, $H$ and $W$ represent the height and width of the image, respectively, while $c$ denotes the channel index.

$$\mu = \frac{1}{H \cdot W} \sum_{h=1}^{H} \sum_{w=1}^{W} x_{mix}(c, h, w) \tag{12}$$

$$\sigma^2 = \frac{1}{H \cdot W} \sum_{h=1}^{H} \sum_{w=1}^{W} (x_{mix}(c, h, w) - \mu)^2 \tag{13}$$

Gaussian noise is then sampled and used to perturb the image's distribution. $\beta$ and $\gamma$ represent perturbed parameters of the image distribution,

$$\beta = \mu + \epsilon_1 \cdot \mu \quad \epsilon_1 \sim N(0,1) \tag{14}$$
$$\gamma = \sigma + \epsilon_2 \cdot \sigma \quad \epsilon_2 \sim N(0,1) \tag{15}$$

This perturbation process enhances the model's robustness by simulating variations in the data distribution, effectively addressing domain shift or unseen distributions during training.

Here, we generate the perturbed image $x_{perturbed}$ through the application of perturbations. If the probability threshold is not met, the original image $x_{mix}$ is used without perturbation, allowing the

$$x_{perturbed} = \frac{x_{mix} - \mu}{\sigma} \cdot \gamma + \beta \tag{16}$$

method to introduce uncertainty into the mixed images in a controlled way. Finally, the mixing operation $mix\_op$ is applied between the previously transformed image $\tilde{x}_{k-1}$ and the perturbed image $x_{perturbed}$, weighted by the mixing ratio $\delta$.

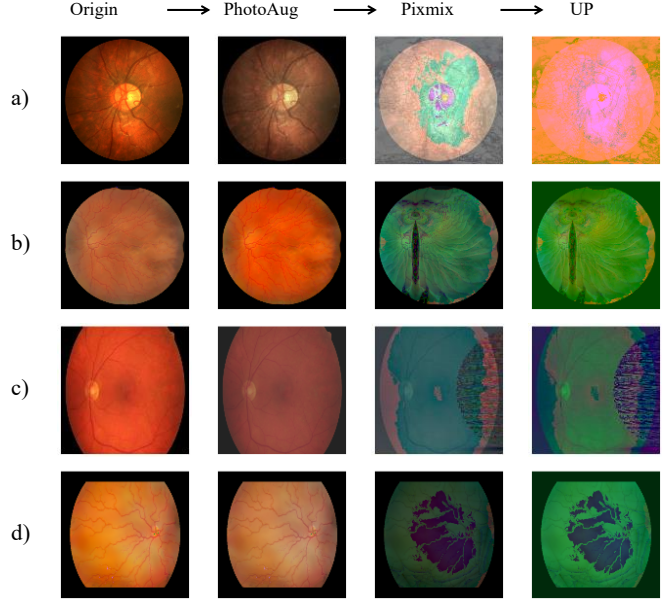

Figure 3 Comparative visualization of style enhancement. The first and third rows illustrate the style augmentation of the original data, while the second and fourth rows depict the style augmentation of the structurally enhanced data.

---

**Algorithm 1** PixMix with PhotoAug and Uncertainty Perturbation

**Input:** Original image $x \in X$, External image $z \in Z$, Maximum mixing rounds $K$, Mixing ratio $\delta \in [0, 1]$, Probability $p \in [0, 1]$
**Output:** Final mixed image $\tilde{x}_K$
1. **Initialize Randomly select initial image augmentation**
2. $\tilde{x}_0$ = random.choice ({photoaug($x$), $x$})
3. **Random choose the number of mixing rounds:**
4. $T$ = random.choice({0, 1, …, $K$})
5. for k = 1 to $T$ do
6. **Sample mixing image:**
7. $x_{mix}$ = random.choice({photoaug($x$), $z$})
8. **Sample mixing operation:**
9. mix_op = random.choice({add, multiply})
10. if random.uniform(0,1) < p then:
11. Apply UP using equation 16
12. end if
13. **Apply mixing operation**
14. $\tilde{x}_k = mix\_op(\tilde{x}_{k-1}, x_{perturbed}, \delta)$
15. end for
16. **Return:** Final mixed image $\tilde{x}_k$

---

Our style enhancement framework enables the generation of more diverse training examples, thereby enhancing the model's robustness to variations in input data. The comprehensive steps of our approach are outlined in Algorithm 2 above.

### 3.4. Overall Loss Function

Our proposed method integrates structural augmentation and stylistic augmentations to enhance the model's ability to generalize across different datasets. The loss function is designed to capture the complexity of both image reconstruction and segmentation tasks. We employ an improved Pix2Pix framework to further train using the vessel-like structures generated by the Space Colonization Algorithm, ensuring that the synthesized retinal vessels closely resemble real retinal images. The generator loss of our improved Pix2Pix model is expressed as follows:

$$L_G = \lambda_{L1}L_{L1} + \lambda_{adv}L_{adv} \tag{17}$$

where the parameter for $\lambda_{L1}$ is set to 100 and the parameter for $\lambda_{adv}$ is set to 0.2. The loss function for the discriminator is defined as follows, with the parameter for $\lambda_1$ set to 0.3 and the parameter for $\lambda_{GP}$ set to 10.

$$L_D = \lambda_1 \left(L_{D_{real}} + L_{D_{fake_{paired}}} + L_{D_{fake_{unpaired}}}\right) + \lambda_{GP} \cdot \left(GP_{paired} + GP_{unpaired}\right) \tag{18}$$

For the segmentation task, we apply a pixel-wise binary cross-entropy loss $L_{BCE}$ to classify each pixel in the image as either part of the target structure (vessel) or background. This loss directly compares the generated segmentation map $x$ with the ground truth map $y$:

$$L_{BCE} = -\mathbb{E}_{x,y}\left[y\log(G(x)) + (1-y)\log(1-G(x))\right] \tag{19}$$

## 4. Experiments

### 4.1. Datasets

We conducted a comprehensive evaluation of the impact of AADG on retinal vessel segmentation based on retinal fundus images. Specifically, experiments were performed on four publicly available datasets: STARE[59], HRF[60], DRIVE[61], and CHASEDB1[62], with sample sizes of 20, 45, 40, and 28, respectively. The images were acquired from various clinical centers, scanners, and populations, ensuring a diverse representation. The partitioning of the training and validation sets adhered to the methodologies established in previous studies[19, 21]. For the domain generalization task, we systematically selected each dataset as the target domain to assess the performance of our method, while utilizing the remaining three datasets as multi-source domains for network training. In this setup, only the training images from the multi-source domains were input into the network, and the network's performance was evaluated exclusively using the test images from the target domain.

### 4.2. Implementation Details

Our DGSSA framework is implemented using the PyTorch framework on an NVIDIA 3090. In the structure enhancement component, we applied a processed Space Colonization Algorithm to generate 100 vessel structure images for each of the four datasets, Subsequently, we employed our improved Pix2Pix model to train on both the vascular structures and the corresponding paired training set to synthesize retinal images. The training process was conducted over 300 epochs with a batch size of 1. We ensured that both unpaired and paired datasets were utilized simultaneously during each epoch. This approach ultimately facilitated the generation of images that closely resemble the original, thereby achieving effective structural enhancement. We employed DeepLabv3+ with a MobileNetv2 backbone as our segmentation models to ensure a fair comparison with other state-of-the-art (SOTA) methods, initializing the backbones with weights from ImageNet. The entire framework was trained using the Adam optimizer. Given the small sample sizes in retinal vessel and lesion segmentation tasks, we set epoch as 300, adjusting the image size to 512x512 pixels. During the training phase, images underwent random scaling, cropping, rotation, and flipping to augment the training dataset. These four types of operations (scaling, cropping, rotation, and flipping) were regarded as default augmentation techniques for the training set, facilitating effective comparisons with previous methodologies. The performance of DGSSA, as well as other SOTA methods, was evaluated using the Dice Similarity Coefficient (DSC) across all segmentation tasks. Additionally, accuracy (ACC), the area under the ROC curve (AUCROC), specificity (SP), RECALL, and PRECISION served as supplementary evaluation metrics for the vessel segmentation task on fundus images. To ensure the reliability and robustness of the experimental results, we validated the experiments through repeated operations.

### 4.3. Comparison with Other Methods

First and foremost, it is important to note that our baseline consists of a model without any generalization techniques, relying solely on default augmentations and a unified segmentor for segmentation. In Table 1, we present an effective comparison of our method against various domain generalization approaches, including M-mixup [63], CutMix [64], DoFE [19], BigAug [57], ELCFS, and AADG [21]. Since our dataset partitioning and training methodology for the domain generalization task align with those of DoFE and AADG, we directly reference the results of DOFE, AADG, and their reproductions with several other methods. Notably, we observe that the style enhancement techniques M-mixup and CutMix, originally applied in natural image classification tasks, exhibit a decline in the Dice Similarity Coefficient (DSC) of 18.9% and 2.1%, respectively, compared to the baseline method. Other domain generalization methods have been effectively utilized in medical image domain generalization tasks, demonstrating improvements over the baseline. Among these, BigAug showed only a modest increase of 0.69% in average DSC, while the ELCFS [20] baseline for vessel segmentation improved by 1.31%. AADG further enhanced performance, achieving a 0.47% increase in DSC for the HRF dataset (referred to as Domain B in Table 1), although the first two methods struggled to generalize effectively on this challenging dataset. AADG exhibited a substantial improvement of 3.29% in average DSC compared to the baseline. By integrating structural enhancement and style enhancement, our approach significantly improves the metrics for retinal vessel segmentation, yielding a 0.60% increase on the HRF dataset. Additionally, compared to AADG, our method achieves a further increase of 3.78% over the baseline, demonstrating the effectiveness of our approach. Moreover, we also achieved notable improvements across the other five evaluation metrics, further highlighting the potential clinical applicability of our method. As illustrated in Figure 4, DGSSA achieves superior preservation of fine vascular structures (e.g., capillary continuity and branching patterns) compared to baseline and AADG.

Table 1 The results for domain a are derived from the model trained with images from the other domains, specifically domains a, b, c, and d corresponding to STARE, HRF, DRIVE, and CHASEDB1, respectively. Top results are highlighted in **bold**, while a dash ('-') indicates that DoFE does not report dice similarity coefficient (DSC) values. This table serves as a comparative analysis against other methods. Additionally, the use of the '-' for certain metrics indicates that these comparative methods follow the same reporting convention as in the AADG paper; therefore, their SP, RECALL, and PRECISION values are not provided here.

| Method | DSC | | | | | AUC-ROC | | | | | ACC | | | | | SP | | | | | RECALL | | | | | PRECISION | | | | |
|---|---|---|---|---|---|---|---|---|---|---|---|---|---|---|---|---|---|---|---|---|---|---|---|---|---|---|---|---|---|---|
| | A | B | C | D | AVG | A | B | C | D | AVG | A | B | C | D | AVG | A | B | C | D | AVG | A | B | C | D | AVG | A | B | C | D | AVG |
| Baseline | 76.32 | 72.23 | 76.27 | 75.71 | 75.13 | 97.36 | 92.18 | 94.50 | 95.97 | 95.00 | 94.10 | 90.46 | 92.07 | 93.45 | 92.52 | 94.81 | 94.24 | 94.92 | 93.30 | 94.32 | 80.12 | 64.54 | 74.83 | 76.45 | 73.98 | 75.41 | 74.37 | 80.25 | 72.8 | 75.71 |
| M-Mixup [63] | 61.89 | 60.41 | 58.22 | 63.21 | 60.93 | 96.02 | 91.76 | 94.30 | 95.45 | 94.38 | 92.32 | 89.25 | 89.05 | 92.46 | 90.77 | - | - | - | - | - | - | - | - | - | - | - | - | - | - | - |
| CutMix [64] | 74.71 | 69.86 | 74.34 | 75.08 | 73.50 | 97.46 | 92.06 | 94.44 | 95.70 | 94.91 | 94.03 | 90.71 | 91.84 | 93.80 | 92.60 | - | - | - | - | - | - | - | - | - | - | - | - | - | - | - |
| BigAug [57] | 79.61 | 70.06 | 76.42 | 76.50 | 75.65 | 97.36 | 90.59 | 94.78 | 96.09 | 94.70 | 94.21 | 89.71 | 91.96 | 93.59 | 92.37 | - | - | - | - | - | - | - | - | - | - | - | - | - | - | - |
| DoFE [19] | - | - | - | - | - | 97.25 | 89.48 | 94.23 | 96.28 | 94.31 | 94.57 | 89.49 | **94.11** | 90.51 | 92.17 | - | - | - | - | - | - | - | - | - | - | - | - | - | - | - |
| ELCFS [20] | 80.92 | 71.85 | 76.61 | 76.40 | 76.44 | 97.82 | 92.18 | 95.14 | 96.29 | 95.36 | 94.54 | 90.57 | 92.27 | 93.44 | 92.71 | - | - | - | - | - | - | - | - | - | - | - | - | - | - | - |
| AADG [21] | 81.79 | 72.57 | 77.70 | 78.34 | 77.60 | **97.96** | 91.93 | 95.21 | **96.95** | 95.51 | 94.75 | 90.49 | 92.41 | 93.84 | 92.87 | **96.89** | 95.56 | 96.31 | 95.17 | 95.98 | 85.10 | **68.20** | 75.10 | 82.80 | 77.80 | 79.10 | **78.81** | 82.30 | 74.20 | 78.60 |
| DGSSA (ours) | **82.17** | **72.66** | **78.62** | **78.45** | **77.98** | 97.95 | **92.51** | **95.70** | 96.90 | **95.77** | **94.77** | **91.00** | 92.52 | **93.98** | **93.07** | 96.27 | **95.88** | **96.62** | **95.91** | **96.17** | **85.34** | 67.94 | **75.35** | **83.00** | **77.91** | **79.35** | 78.21 | **82.51** | **74.48** | **78.64** |

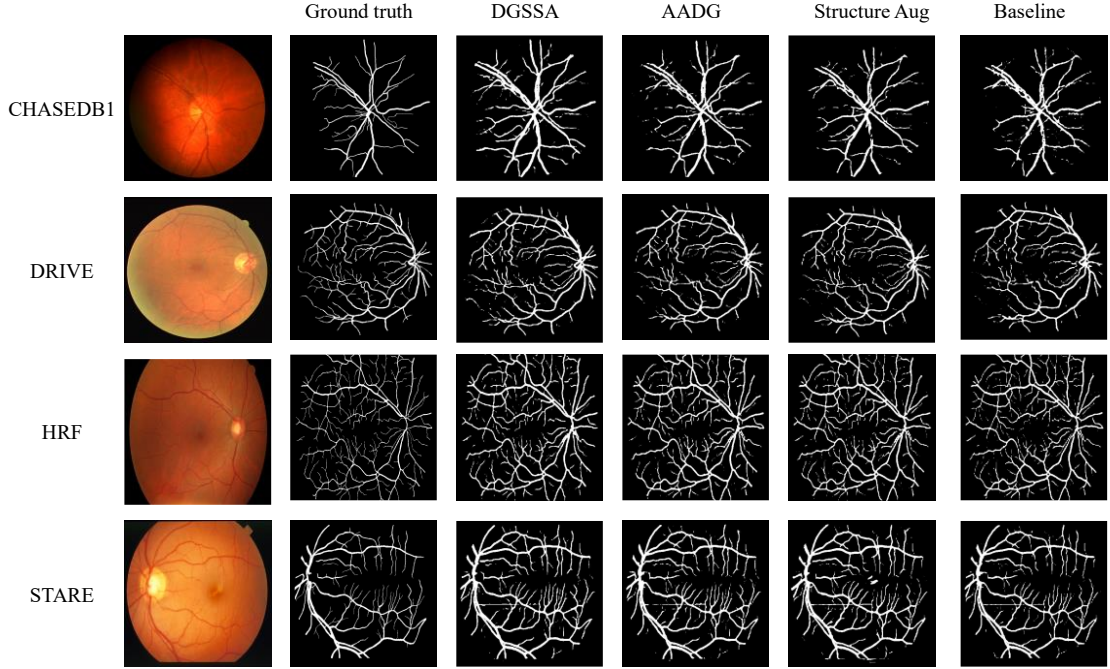

Figure 4 Comparison of segmentation results among DGSSA, Structure Augmentation, AADG, and the baseline method.

## 4.4. Ablation Study of DGSSA

In Table 2, our structural augmentation significantly surpasses the baseline, highlighting its effectiveness. Furthermore, we conducted ablation experiments that utilized only stylistic augmentation and those that excluded spatial perturbations. As illustrated in the table, the implementation of structural augmentation alone results in improvements over the baseline, and when combined with stylistic augmentation, further enhancements are achieved. By eliminating the influence of spatial perturbations, we observe that the DSC for our method does not significantly differ from the training approach without Uncertainty Perturbation (UP). However, we achieve superior performance in the AUC-ROC, accuracy (ACC), SP, RECALL and PRECISION metrics, indicating that spatial perturbations play a role in enhancing overall performance. These further validate the necessity of our combined approach to optimization through both structural and stylistic enhancements.

Table 2 These are the results of our ablation studies, which include experiments focused solely on structural augmentation, solely on stylistic augmentation, and an additional experiment without uncertainty perturbations.

| Method | DSC | | | | | AUC-ROC | | | | | ACC | | | | | SP | | | | | RECALL | | | | | PRECISION | | | | |
|---|---|---|---|---|---|---|---|---|---|---|---|---|---|---|---|---|---|---|---|---|---|---|---|---|---|---|---|---|---|---|
| | A | B | C | D | AVG | A | B | C | D | AVG | A | B | C | D | AVG | A | B | C | D | AVG | A | B | C | D | AVG | A | B | C | D | AVG |
| Baseline | 76.32 | 72.23 | 76.27 | 75.71 | 75.13 | 97.36 | 92.18 | 94.5 | 95.97 | 95.00 | 94.10 | 90.46 | 92.07 | 93.45 | 92.52 | 94.81 | 94.24 | 94.92 | 93.30 | 94.32 | 80.12 | 64.54 | 74.83 | 76.45 | 73.99 | 75.41 | 74.37 | 80.25 | 72.80 | 75.71 |
| Structure Augmentation | 79.47 | 72.55 | 77.97 | 77.35 | 76.83 | 97.41 | 92.48 | 94.25 | 96.12 | 95.06 | 94.39 | 90.73 | 92.39 | 93.82 | 92.83 | 95.55 | 94.68 | 96.84 | 94.17 | 95.31 | 80.39 | 65.94 | 73.40 | 77.88 | 74.41 | 74.41 | 79.81 | 74.99 | 81.02 | 74.83 | 77.66 |
| Style Augmentation | 81.56 | 72.44 | 77.81 | 77.95 | 77.44 | 97.75 | 92.11 | 95.21 | 96.76 | 95.46 | 94.65 | 90.90 | 92.38 | 93.80 | 92.93 | 95.43 | 95.71 | 96.29 | 95.79 | 95.81 | 84.35 | 67.81 | 70.78 | 81.71 | 76.16 | 80.01 | 77.04 | 81.44 | 74.65 | 78.23 |
| DGSSA without UP | 82.24 | 72.69 | 78.54 | 78.39 | 77.96 | 97.99 | 92.08 | 95.32 | 96.90 | 95.57 | 94.84 | 90.93 | 92.34 | 93.81 | 92.98 | 96.35 | 95.21 | 95.88 | 95.71 | 95.79 | 83.00 | 65.95 | 75.29 | 80.97 | 76.30 | 80.84 | 78.10 | 80.66 | 74.46 | 78.52 |
| DGSSA | 82.17 | 72.66 | 78.62 | 78.45 | 77.98 | 97.95 | 92.51 | 95.70 | 96.90 | 95.77 | 94.77 | 91.00 | 92.52 | 93.98 | 93.07 | 96.27 | 95.88 | 96.62 | 95.91 | 96.17 | 85.34 | 67.94 | 75.35 | 83.00 | 77.91 | 79.35 | 78.21 | 82.51 | 74.48 | 78.64 |

To further validate our experiments, we conducted three additional trials using U-Net as our segmentation network, using Resnet as backbone and maintaining initialization with ImageNet weights, as shown in Table 3. Table 3 demonstrates that our method generalizes effectively across various segmentation architectures, underscoring its versatility and robustness. Note that using U-net as the segmentation network can achieve better performance with a dice score of 84.75, further surpassing existing methods. This adaptability highlights the potential of our approach to enhance performance not only within our primary framework but also across different models, thereby contributing to broader applications in medical image segmentation tasks.

Table 3 DSC metrics for segmentation ablation experiments using U-net and Deeplabv3+ on the STARE dataset.

| Network | Baseline | Structure Augmentation | DGSSA |
| --- | --- | --- | --- |
| Deeplabv3+ | 76.32 | 79.47 | 82.17 |
| U-net | 81.23 | 82.12 | 84.75 |

### 4.5. Hyperparameters Sensitivity Analysis

To ensure the stability and reliability of our method, we conducted a sensitivity analysis on key hyperparameters. This experiment aims to evaluate how variations in loss term weights affect segmentation performance. As illustrated in the Figure 5, the Dice coefficient remains stable across a broad range of values for $\lambda_1$, $\lambda_{adv}$, and $\lambda_{L1}$, demonstrating the robustness of our method to these hyperparameters. In contrast, increasing $\lambda_{GP}$ results in a decline in performance, suggesting that overemphasis on the gradient penalty term may negatively impact optimization. Notably, across all tested ranges, our method consistently outperforms AADG [21], further highlighting its effectiveness and resilience to hyperparameter fluctuations.

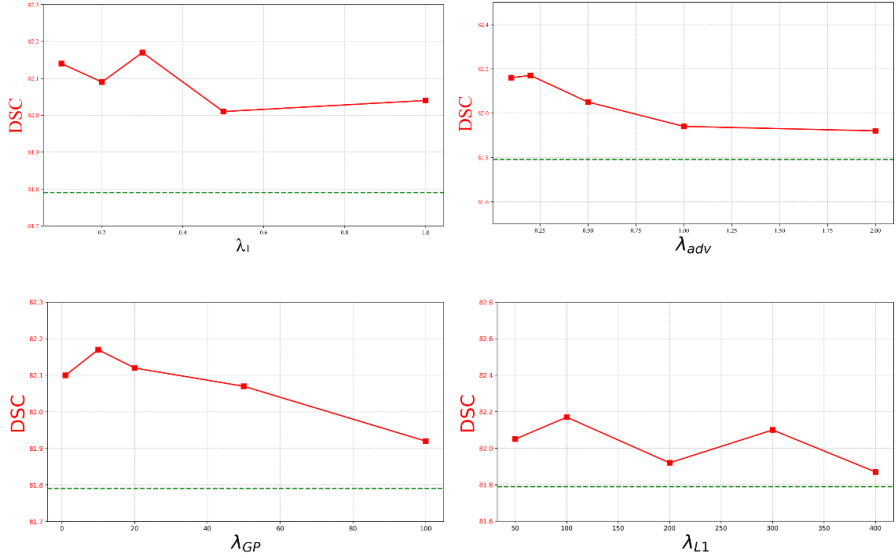

Figure 5 Sensitivity analysis with respect to hyperparameters $\lambda_1$, $\lambda_{adv}$, $\lambda_{GP}$ and $\lambda_{L1}$. The green dashed line represents the average Dice score of the AADG segmentation method, while the red line plot represents the performance of our method in retinal vessel segmentation under different parameter settings.

## 5. Discussion

**5.1 Comparison with Existing Domain Generalization Methods**

As show in Table 1 and Figure 4, our experiments demonstrate that the DGSSA framework significantly improves the performance of retinal vessel segmentation compared to existing domain generalization methods. In prior approaches, such as DoFE, domain information is aggregated into a single feature space; however, this method does not generalize well for retinal vessel segmentation. BigAug employs manually defined parameters to apply a set of image transformations. Although it exhibits some improvements in domain generalization tasks, its conservative design limits the range of transformations (e.g., adjusting brightness by randomly varying intensity within the range of [-0.1, 0.1]), resulting in insufficient enhancement of distribution diversity. ELCFS shows a 1.31% increase in the DSC for vessel segmentation, attributable to its frequency-space interpolation mechanism, which continuously generates images exhibiting features from other domains. AADG utilizes adversarial training and deep reinforcement learning to effectively search for targets, thereby streamlining the automatic enhancement process. While it represents an improvement over previous metrics, it is still optimized solely for stylistic enhancement, indicating a need for further augmentation. Consequently, we adopt a novel perspective on optimizing domain generalization by effectively combining structural and stylistic enhancements. Our results confirm that this approach is indeed effective.

**5.2 Segmentation Error Analyses**

In Figure 4, we conducted a qualitative analysis of the segmentation results, which clearly demonstrates that DGSSA significantly enhances segmentation performance. In Figure 6, we present a comprehensive quantitative error analysis by visualizing the confusion matrices and segmentation results for representative samples from four datasets. This figure not only highlights the segmentation accuracy of DGSSA but also enables a deeper understanding of the spatial distribution of errors. The overlay visualizations reveal that most false positives and false negatives are concentrated around thin vessel branches and vessel boundaries, where segmentation is inherently more challenging. Combined with Figure 4, it is evident that our method achieves further improvements in capturing fine vessel branches. We obtained promising results on the CHASEDB1, DRIVE, and STARE datasets; however, there remain many areas requiring further refinement. In particular, the HRF dataset—known for its high complexity and the presence of numerous micro-vessels—poses a significant challenge. This is also reflected in the confusion matrix, where a relatively large number of false negatives suggests the model struggles to capture fine vascular details. These intricate structures highlight the need for continued improvement in future work.

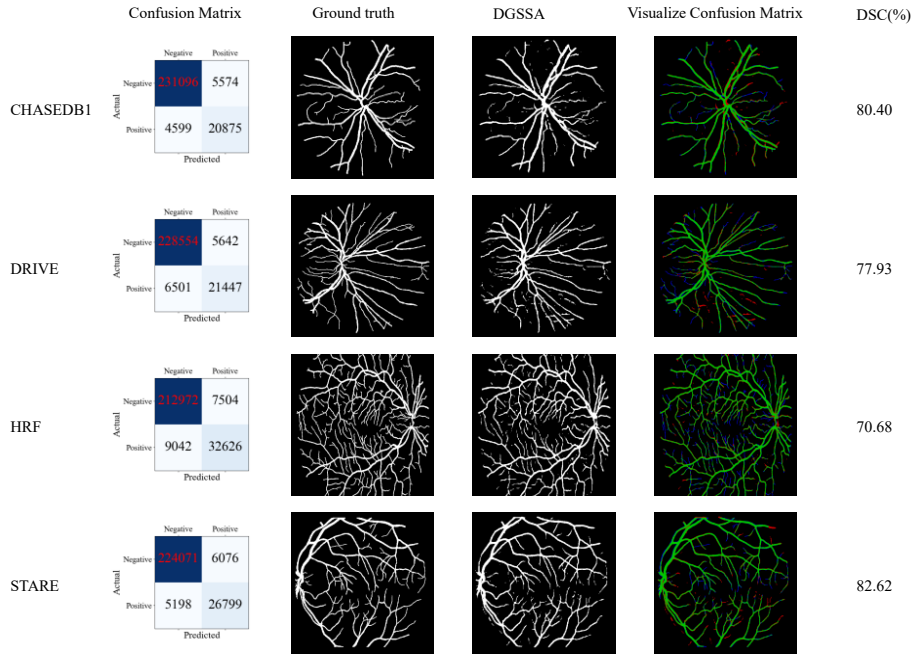

Figure 6 Visualization of the confusion matrix for individual samples. Light green indicates true positives (TP), light red denotes false positives (FP), and light blue represents false negatives (FN). Rows 1 to 4 correspond to samples from the CHASEDB1, DRIVE, HRF, and STARE datasets, respectively.

**5.3 Qualitative and Quantitative Analysis of Domain Generalization Performance**

Furthermore, in Figure 7, we present a t-SNE dimensionality reduction visualization of the distributions resulting from various augmentation strategies. The visual representation illustrates that our structural augmentation effectively enhances the sample size and, when combined with stylistic augmentation, further increases the diversity of the samples, qualitatively demonstrating the significant impact of our methodology. This enhancement not only increases the diversity of the training data but also facilitates better generalization capabilities of the segmentation model across different datasets, thus reinforcing the advantages of our integrated approach.

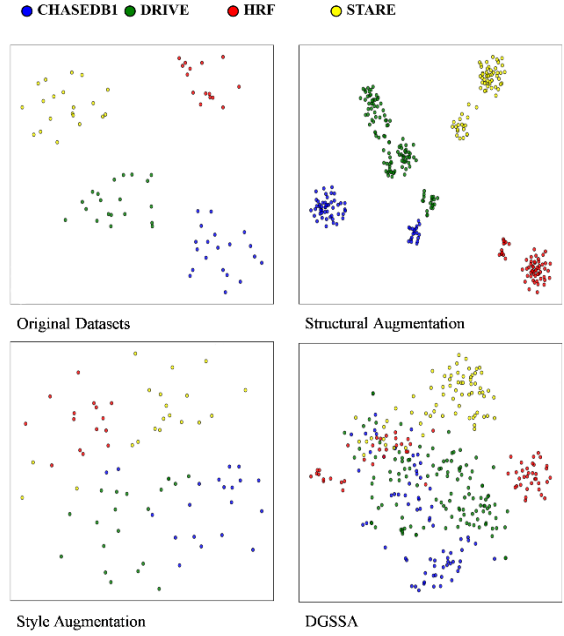

Figure 7 t-SNE visualization of VGG16 features extracted from augmented retinal images, with four colors representing the four training datasets.

Table 4 Quantitatively reflects the degree of domain separation by calculating the inter distances between domain centers using Euclidean distance.

|  | Original Dataset | Structural Augmentation | Style Augmentation | DGSSA |
| --- | --- | --- | --- | --- |
| Inter Distance | 78.15 | 72.95 | 61.04 | 52.20 |

Additionally, we quantitatively calculated the domain distance, which represents the differences between the centers of different domains and reflects the degree of separation between domains. From Table 4, it can be seen that as our method improves, the degree of domain separation decreases, which is consistent with the results shown by t-SNE, further validating the reliability of our DGSSA method.

## 5.4 Statistical Comparison

To further demonstrate the effectiveness of our approach, we conducted a statistical test (paired t-test) and employed a Critical Difference (CD) diagram [65] to compare our method with other techniques. From these results in Figure 8 and Table 5, we observe that, compared to the version without Uncertainty Perturbation (UP), our method does not exhibit a significant advantage on the DSC metric; however, it achieves superior performance across the remaining metrics. Moreover, when compared with alternative state-of-the-art methods and other ablation variants, our approach consistently delivers the best results and demonstrates statistically significant improvements.

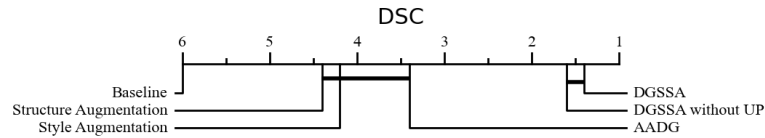

Figure 8 A statistical comparison on the DSC metric was conducted using a Critical Difference (CD) diagram to evaluate our method against other approaches as well as ablation studies.

Table 5 A paired t-test was performed at the sample level to evaluate the performance of our method, along with other approaches and ablation experiments, on the retinal vessel segmentation task.

| Method | DGSSA without UP | AADG [21] | Style Augmentation | Structure Augmentation | Baseline |
|---|---|---|---|---|---|
| DGSSA(Ours) | 0.31 | 0.0488 | 0.032 | 0.019 | <0.01 |

## 5.5 Quantitative Analysis on Complex Vessel Structures

To quantitatively validate the model's performance on complex vessel regions, we conducted a targeted evaluation that isolates structures of varying morphology. In line with our error analysis (Section 5.2), which identifies thin vessel branches as a primary segmentation challenge, this evaluation partitions vessel structures based on their diameter. For both the ground truth ($T$) and the model's prediction ($P$), we independently classified all vessel pixels into "thin" ($T_{thin}$, $P_{thin}$) and "thick" ($T_{thick}$, $P_{thick}$) subsets. This was achieved by estimating the vessel radius at each point along its skeletonized centerline, using a threshold of 1.2 pixels to differentiate thin structures.

The performance on these specific regions was then assessed by comparing the corresponding partitions. For example, the DSC for thin vessels was calculated as:

$$DSC_{thin} = \frac{2 \cdot |P_{thin} \cap T_{thin}|}{|P_{thin}| + |T_{thin}|} \quad (20)$$

The calculation for thick vessels follows the same principle. The results of this targeted evaluation are presented in Table 6. This analysis is intentionally stringent, as it penalizes not only detection errors but also morphological inaccuracies (e.g., predicting a thin vessel with an incorrect diameter). This strictness results in lower absolute DSC scores compared to standard overlap-based metrics. The key finding, however, lies not in the absolute scores but in the relative performance gains. The data in Table 6 offers a clear deconstruction of how each component contributes to performance. The analysis reveals that Structure Augmentation is the principal factor in improving thin-vessel segmentation, elevating the average DSC from a baseline of 31.79% to 35.23%. This gain is notably more significant than that provided by Style Augmentation (35.02%), confirming that directly enriching morphological variety is more effective for this specific challenge than diversifying styles alone. However, the optimal performance is achieved when both strategies are synergistically combined within our full DGSSA framework, which reaches the highest average thin-vessel DSC of 35.95%. This represents a 13.1% relative improvement over the baseline for thin vessels, a figure that stands in stark contrast to the modest 2.8% gain on thick vessels. This disparity provides compelling quantitative evidence that our framework's effectiveness on complex regions is driven primarily by structural enhancements, which are then amplified by stylistic variations to achieve state-of-the-art performance.

Table 6 DSC performance on thin and thick vessel structures across all datasets. Datasets A, B, C, and D correspond to STARE, HRF, DRIVE, and CHASEDB1, respectively.

| Method | Thin Vessel DSC | | | | | Thick Vessel DSC | | | | |
|---|---|---|---|---|---|---|---|---|---|---|
| | A | B | C | D | AVG | A | B | C | D | AVG |
| baseline | 33.24 | 34.22 | 32.31 | 27.40 | 31.79 | 73.75 | 66.68 | 72.98 | 69.28 | 70.67 |
| Structure Augmentation | 39.57 | 36.96 | 35.03 | 29.37 | 35.23 | 76.16 | 67.35 | 73.44 | 70.95 | 71.98 |
| Style Augmentation | 39.91 | 36.28 | 34.62 | 29.28 | 35.02 | 76.41 | 67.40 | 73.39 | 71.16 | 72.09 |
| DGSSA | 41.20 | 37.15 | 35.84 | 29.59 | 35.95 | 76.76 | 67.75 | 73.90 | 72.12 | 72.63 |

## 5.6 Computational Complexity

The computational overhead of our method arises primarily during the data generation and training phases, while the inference phase remains unaffected. During data preparation, the main computational cost is attributed to training the retinal image generation network based on the pix2pix architecture. This network is used to synthesize anatomically plausible retinal images guided by vascular structures. For each dataset, training the pix2pix model typically takes approximately 100 minutes, as it incorporates both real and synthetically generated vessel structure maps during training. Once trained, the image synthesis process is computationally lightweight: generating the full set of augmented samples for a dataset takes only 1–2 minutes, given the relatively small size of the generation model. In the training phase of the segmentation model, stylistic augmentation is performed at the image level using preprocessed style-transferred images. These are fused with the original images before training begins, and thus do not introduce additional processing during training iterations. The primary training overhead stems from the structural augmentation component, which increases the number of training samples. This results in a 2–3× increase in training time compared to the baseline model, depending on the volume of synthetic samples incorporated. Importantly, our method introduces no additional computational burden during inference. We do not modify the architecture of the segmentation model or append any auxiliary modules; thus, inference speed and memory usage remain identical to those of the baseline models. This characteristic makes our framework particularly well-suited for clinical applications, where real-time or near real-time performance is often essential.

**5.7 Clinical Implications**

The proposed DGSSA framework holds considerable promise for integration into real-world clinical workflows. Its design ensures that no additional network modules are needed, and it imposes no extra computational burden during inference, making it particularly well-suited for deployment in clinical settings, including those with resource constraints. In such environments, real-time efficiency and system compatibility are critical. Furthermore, our structural augmentation technique plays a crucial role in addressing the challenge of data scarcity, a common issue in medical image analysis where annotated datasets are often limited. By generating high-quality synthetic samples with diverse anatomical and stylistic variations, we enhance the robustness of segmentation models, improving their ability to generalize across different patient populations and imaging conditions.

The core motivation for our structural augmentation technique stems from the recognition that retinal vessel structures exhibit inherent variability across patients. This variability often poses challenges for segmentation models, particularly in clinical settings where new patient cases, each with unique anatomical features, are encountered daily. Our augmentation approach helps address this issue by creating synthetic samples that represent a wider spectrum of retinal vessel structures, ultimately improving segmentation accuracy and consistency. This is particularly important in clinical practice, where models must perform reliably across diverse patient populations and imaging conditions. By enhancing model robustness and generalization, our framework can significantly improve the overall quality of retinal vessel segmentation, supporting clinicians in making more accurate and timely diagnoses.

Our framework's potential extends beyond retinal vessel segmentation. The combination of structural and stylistic augmentation methods holds promise for a wide array of medical image analysis tasks, such as tumor segmentation and other diagnostic imaging challenges, where data scarcity and domain shift are common issues. By addressing these challenges, the DGSSA framework has the potential to transform clinical workflows, providing clinicians with a powerful tool that can assist in the accurate and efficient segmentation of medical images.

**5.8 Limitations and Future Improvements**

The effectiveness of our method has been particularly evident in the segmentation of tubular structures, such as blood vessels, suggesting its potential within similar contexts in medical image analysis. A key element driving the success of our approach is the integration of both structural and stylistic augmentation. Structural augmentation enables the generation of realistic vascular structures, expanding the training dataset and providing a more comprehensive representation of vascular morphology. This is crucial for tubular structures, which exhibit significant variation in diameter, branching patterns, and curvature. While the structural augmentation technique employed in this study is specifically tailored for vessel-like anatomical structures, the stylistic augmentation strategy is general in nature and can be readily extended to other types of organs and segmentation tasks. This distinction underscores both the strengths and current limitations of our method, particularly in its adaptability to broader medical analysis applications.

The datasets employed in this study—STARE, HRF, DRIVE, and CHASEDB1—each have unique characteristics that significantly influence model performance. For example, STARE offers high-resolution images with clearly defined vessels, which contribute to more accurate segmentation. In contrast, CHASEDB1 poses challenges due to lower image quality and lighting variability, potentially hindering the model's ability to generalize. Furthermore, the relatively small sample sizes of these datasets may limit the generalizability of the model to larger, real-world clinical environments, thereby increasing the risk of overfitting. Our proposed structural and stylistic augmentation approach helps mitigate this data scarcity to some extent by generating high-quality synthetic samples that enhance diversity and robustness.

To address these challenges, future work should focus on developing more generalizable structural augmentation strategies that are not limited to vascular anatomy. Although our current implementation is optimized for vessel-like structures, advancing the structural augmentation component to accommodate a wider range of anatomical forms will be essential for improving the flexibility and applicability of this method. Additionally, dataset-specific preprocessing strategies could be explored; for instance, in liver segmentation, label deformation could be applied before synthesizing high-quality images for data augmentation. In our case, we employed the spatial colonization algorithm due to its ability to generate realistic vascular structures. Looking forward, more advanced techniques, such as diffusion models, could be incorporated to achieve even higher fidelity in structural augmentation. These efforts will be essential for developing versatile and robust segmentation models capable of addressing a wider spectrum of clinical tasks.

In parallel, future work could also strengthen the clinical applicability and generalization ability of the model by incorporating additional, larger-scale datasets from diverse imaging devices and patient populations. These efforts together will be essential for developing versatile and robust segmentation models capable of addressing a wider spectrum of clinical tasks.

## 6. Conclusion

In this paper, we introduce a novel domain generalization method for retinal vessel image segmentation that effectively integrates structural and stylistic enhancement strategies. Through a rigorous evaluation conducted on four challenging retinal vessel segmentation datasets—STARE, HRF, DRIVE, and CHASEDB1—we demonstrate that our method achieves notable improvements in performance, with Dice Similarity Coefficients (DSC) of 82.17%, 72.66%, 78.62%, and 78.45%, respectively. Our approach achieves state-of-the-art results, surpassing existing methods. These findings validate the efficacy of our proposed framework and highlight the importance of combining structural and stylistic enhancements to improve segmentation tasks. Our research offers promising avenues for further exploration of domain generalization in the context of medical image analysis.

**Declaration of competing interest**

The authors declare that there are no conflicts of interest regarding the publication of this paper.

**CRediT authorship contribution statement**

**Bo Liu**: Conceptualization; Data curation; Formal analysis; Investigation; Methodology; Software; Validation; Visualization; Writing-original draft. **Yudong Zhang**: Validation, Writing-review & editing. **Shuihua Wang**: Data curation, Writing-review & editing. **Siyue Li**: Data curation, Writing-review & editing. **Jin Hong**: Conceptualization; Data curation; Investigation; Methodology; Resources; Funding acquisition; Project administration; Supervision; Writing-review & editing.

**Acknowledgements**

This work was supported in part by the National Natural Science Foundation of China (62466033), in part by the Jiangxi Provincial Natural Science Foundation (20242BAB20070).